# Fluctuation conductivity and pseudogap in optimally doped $SmFeAsO_{1-x}$

A.L. Solovjov[1]*, S.L. Sidorov[2], Yu.V. Tarenkov[2] and A.I. D'yachenko[2]

[1] *B.Verkin Institute for Low Temperature Physics & Engineering National Academy of Sciences of Ukraine, Lenin Ave. 47, 61103 Kharkov, Ukraine*
[2] *A. Galkin Institute for Physics & Engineering National Academy of Sciences of Ukraine, R .Luxemburg 72, 83114 Donetzk, Ukraine*

We report the analysis of fluctuation conductivity (FLC) and pseudogap ($\Delta^*$) in Fe-based superconductor $SmFeAsO_{0.85}$ with $T_c$ = 55 K derived from resistivity experiments. In the FLC analysis distinct 2D-3D (MT-AL) crossover typical for cuprates but followed by enhanced MT fluctuation contribution was found. Using the crossover temperature $T_0$ coherence length along c-axis $\xi_c(0)$ was determined. Rather specific $\Delta^*(T)$ dependence with two representative temperatures followed by minimum at about 125 K was observed. Below $T_s \approx$ 147 K, corresponding to SDW ordering, $\Delta^*(T)$ decreases linearly down to $T_{AF} \approx$133 K. This last peculiarity can likely be attributed to antiferromagnetic (AF) ordering in FeAs planes. It is believed that found behavior is strongly associated with specific electronic configuration of the Fe-based compounds.

## I. INTRODUCTION

Recent discovery of a new class of high-temperature superconductors (HTS's) with an active *As-Fe-As* plane [1], in which $T_c$ ranges from 26 K in $LaO_{1-x}F_xFeAs$ [1] up to 55 K in $SaFeAsO_{0.85}$ [2], has evidently regained interest to the problem of high-temperature superconductivity. The discovery stimulated appearance of a huge amount of papers in which the main features of electronic spectrum, taking into consideration the role of correlations and collective excitations (phonons, spin waves etc.) as well as possible mechanisms of magnetic ordering and Cooper pairing in these compounds, are studied [3-5]. A comparison of results of these experiments with those obtained for high-$T_c$ cuprate superconductors is to shed a light on a problem of pairing mechanism and pseudogap (PG) phenomenon in both superconducting systems and likely to clarify the nature of the high-temperature superconductivity on the whole. However, no results as for fluctuation conductivity (FLC) (sometimes regarded as paraconductivity [6]) are reported so far. Besides, there is an evident lack of experimental results as for pseudogap (PG) measurements. In this paper the analysis of FLC and PG derived from resistivity measurements of $SmFeAsO_{1-x}$ high-$T_c$ superconducting system based on the model developed in our previous papers [7,8] is presented.

In resistivity measurements of cuprates, e.g. in YBCO, the PG demonstrates itself as a pronounced downturn of the longitudinal resistivity $\rho_{xx}(T)$ at $T \leq T^*$ from its linear dependence at higher temperatures. This results in appearance of the excess conductivity $\sigma'(T) = \sigma(T) - \sigma_N(T)$ which can be written as

$$\sigma'(T)=[\rho_N(T)-\rho(T)]/[\rho_N(T)\rho(T)]. \qquad (1)$$

Here $\rho = \rho_{xx}(T)$ is the measured resistivity, and $\rho_N(T) = aT + b$ determines the resistivity of a sample in the normal state extrapolated to low temperatures. This way of determining $\rho_N(T)$, which is widely used for calculating $\sigma'(T)$ in HTS's [7,8], has been justified by the NAFL model [9].

The Aslamazov-Larkin (AL) [10] and Hikami-Larkin (HL) [11] fluctuation theories describe the excess conductivity $\sigma'(T)$ in YBCO in the temperature interval from $T_c$ up to ~110 K [8] suggesting the presence of fluctuating Cooper pairs in HTS's in the indicated temperature region. The question of whether or not paired fermions may form in HTS's in the whole PG temperature region still remains very controversial. Indeed, it seems unlikely that Cooper pairs satisfying the BCS–Bogolyubov theory are formed at temperatures $T >> 100$ K especially considering the fact that the coherence length in HTS's is extremely short ($\xi_{ab}(0)$ = (10÷15) Å). According to the concept of the local pairs [12,13] such pairs in HTS's could be not interacting with one another strongly bound bosons (SBB) satisfying the Bose-Einstein condensation (BEC) statistics [7,8].

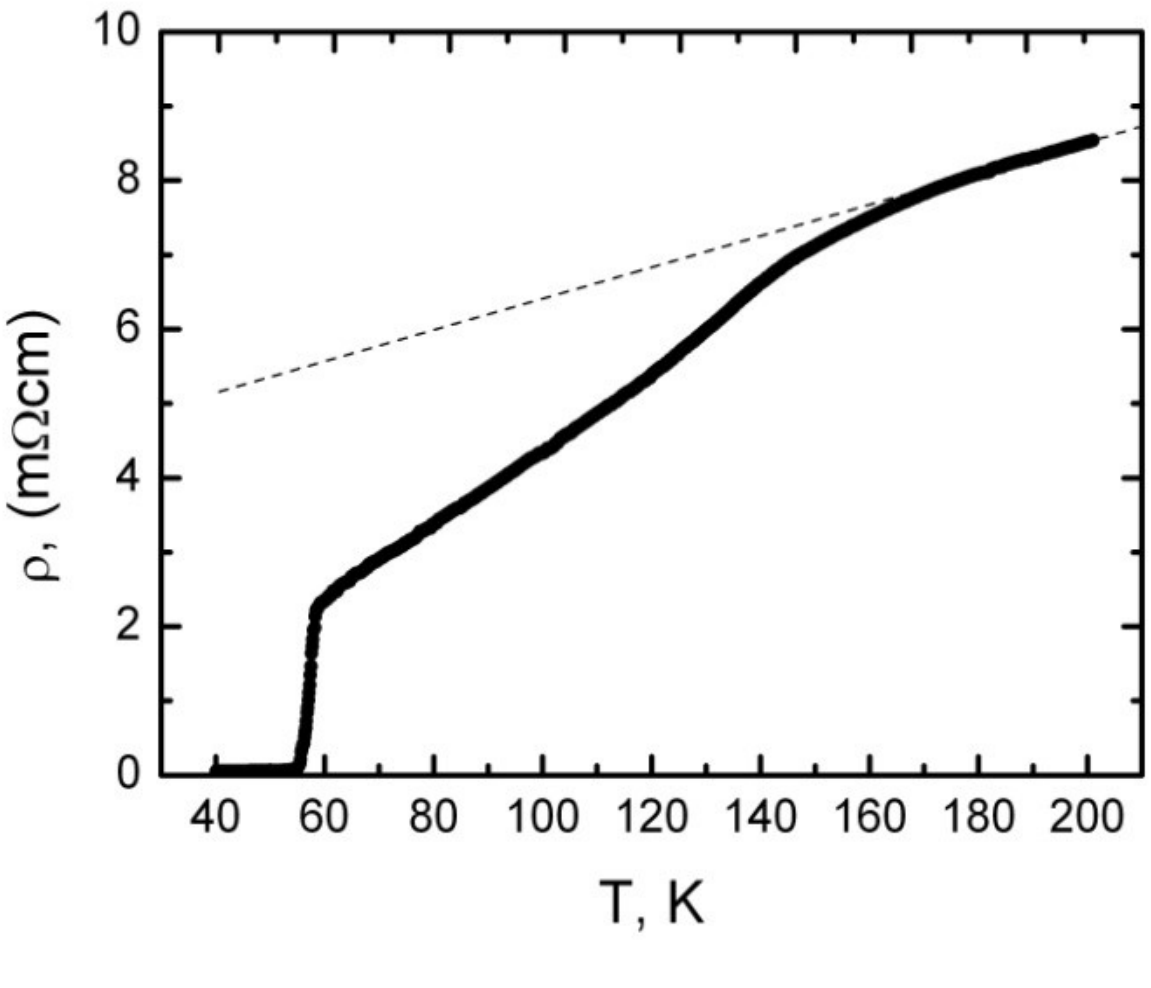

FIG.1: Resistivity $\rho$ versus temperature $T$ (●); dashed line represents $\rho_N(T)$.

In accordance with the theory [13-17], the stable, strongly bound bosons which satisfy the BEC theory can be realized only in systems with low and intermediate charge-carrier density $n_f$. This is just the case for cuprate HTS's [14,18] and new iron-based superconductors [3-5]. Apart from high $T_c$ and PG the low charge-carrier density can be considered as the third fundamental property which distinguishes HTS's from conventional low-temperature superconductors. Existence of SBB in PG state of cuprates as well as BEC-BCS transition upon temperature diminution were experimentally justified in studying YBCO films with different oxygen concentration [7,8]. We expected to find the similar features of the FLC and PG behaviour in iron-based HTS's at least in compounds with currently highest $T_c$ such as $SmFeAsO_{1-x}$.

## II. EXPERIMENTAL DETAILS

Analysis of the FLC and PG in the FeAs compounds was performed using results of resistivity measurements of $SmFeAsO_{0.85}$ polycrystal carried out using fully computerized set up. The sample width and thickness are w=1.5 mm, $d_0$=1.4 mm, respectively. The distance between potential contacts denotes the length of the sample $L$ = 4 mm. Temperature dependence of resistivity $\rho(T)$ shown in Fig.1 (dots). The critical temperature $T_c \cong 55$ K suggests the oxygen index of studied polycrystal to be $x\sim 0.85$. The width of the resistive transition into superconducting (SC) state is $\Delta T \leq 2$ K suggesting good phase and structural uniformity of the sample. In accordance with our approach resistivity curve above ~ 170 K is extrapolated by straight line (dashed line in the figure). Unfortunately, the whole resistivity curve was measured starting from 200 K only. Nevertheless the linear part of the $\rho(T)$ behavior is distinctly seen on the plot down to $T^*$ = (175 ± 3) K allowing us a possibility to get $\rho_N(T)$ and extract the proper values of $\sigma'(T)$ using Eq. (1). In accordance with our verification, the alteration of the slope of the $\rho_N(T)$ dependence in reasonable limits does not affect the temperature dependence of both $\sigma'(T)$ and $\Delta^*(T)$. Nevertheless, to be more confident the resistivity data, reported by Ren et al. [19] for $SmFeAsO_{0.9}F_{0.1}$ with $T_c$ =52.6 K, have been analyzed and very similar slope of the linear $\rho_N(T)$ was found.

## III. RESULTS AND DISCUSSION

### A. Fluctuation conductivity

$\sigma'^{-2}$ as a function of temperature is plotted in Fig. 2 (squares). Interval of it's linear behaviour, fitted by the dashed line, corresponds to temperature region of *3D* AL fluctuations [7,11]. It's crossing with $T$-axis denotes mean-field critical temperature $T_c^{mf} \approx 57$ K. Above $T_0 \approx 58.5$ K (Fig. 2) $\sigma'^{-2}(T)$ apparently deviates toward high temperatures suggesting the presence of Maki-Thompson (MT) fluctuation contribution in the sample. Besides, $T_c^{mf}$ is obviously larger than $T_c$ i.e. lies out of temperature region of critical fluctuations in a good agreement with our approach [8,20].

Taking found value of $T_c^{mf}$ into account, $\sigma'$ as a function of the reduced temperature $\varepsilon = \ln (T/T_c^{mf}) \approx (T- T_c^{mf})/T_c^{mf}$ can be computed. The $\ln\sigma'$ vs $\ln\varepsilon$ is displayed in Fig. 3 at temperature interval relatively close to $T_c$ (dots) and compared with the HL theory in the clean limit (curves 1-3). Up to $T_0 \approx 58.5$ K ($\ln\varepsilon_0 \approx -3.6$) the data are well extrapolated by AL fluctuation contribution for any 3D system (straight line 2 in the figure)

$$\sigma'_{AL}=\{e^2/[32\hbar\xi_c(0)]\}\varepsilon^{-1/2}\,, \quad (2)$$

but above $T_0$, up to $T_{c0} \approx 69$ K ($\ln\varepsilon_{c0} \cong -1.55$), this is 2D MT term of the HL theory (curve 3)

$$\sigma'_{MT}=\{[e^2/[8\hbar d(1-\alpha/\delta)]\}\cdot \cdot\ln\{(\delta/\alpha)[1+\alpha+(1+2\alpha)^{1/2}]/[1+\delta+(1+2\delta)^{1/2}]\}\varepsilon^{-1}, \quad (3)$$

that dominates well above $T_c$ in the 2D fluctuation region [11]. Here $\xi_c(0)$ is the coherence length along the c-axis, i.e. perpendicular to the conducting planes, d = (3.05 ± 0.5) Å is the distance between conducting layers in $SmFeAsO_{1-x}$, $\alpha = 2\xi_c^2(T)/d^2 = 2[\xi_c(0)/d]^2\varepsilon^{-1}$ is the coupling parameter,

$$\delta=1{,}203(l/\xi_{ab})(16/\pi\hbar)[\xi_c(0)/d]^2k_BT\tau_\varphi \quad (4)$$

is the pair-breaking parameter, and $\tau_\varphi$ is the phase relaxation time (lifetime) of the fluctuating pairs. The factor $1.203(l/\xi_{ab})$, where $l$ is the mean-free path and $\xi_{ab}$ is the coherence length in the *ab* plane, takes account of the approach to the clean limit [8,20].

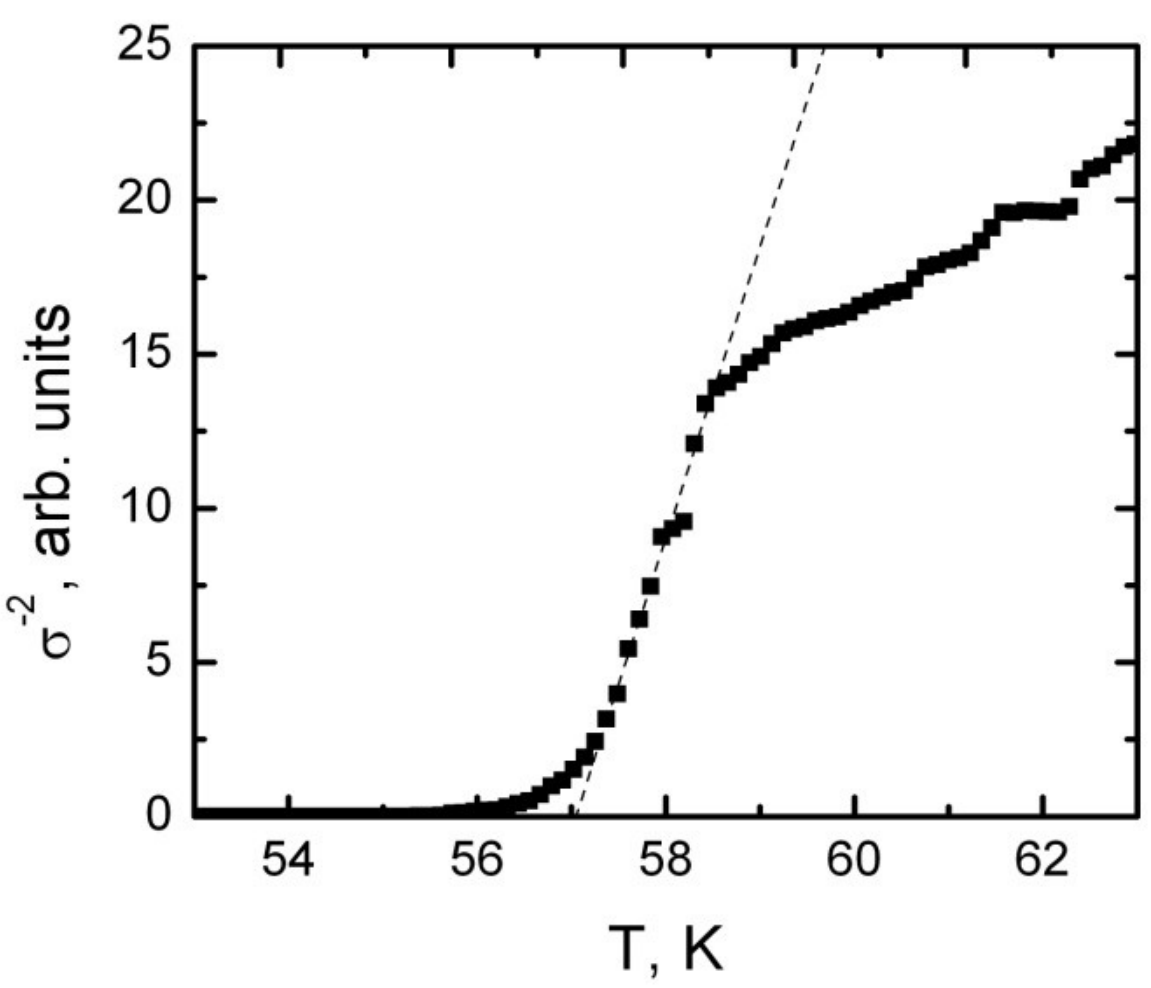

FIG. 2: $\sigma'^{-2}(T)$ (squares) which denotes $T_c^{mf} \approx 57$ K. The dashed line is the guidance for eyes only.

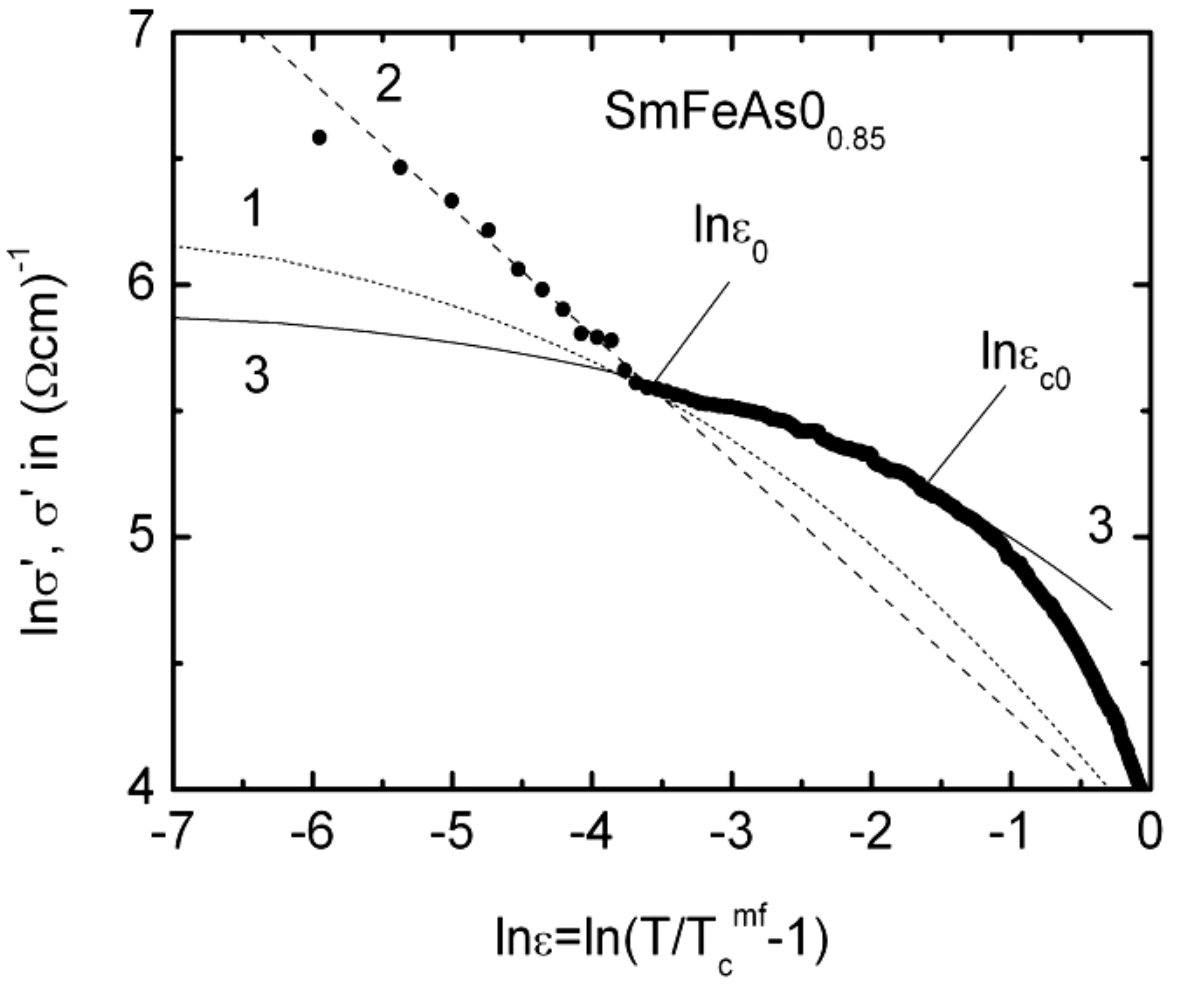

FIG. 3: $ln\sigma'$ as a function of $ln\varepsilon$ near $T_c$ (dots) in comparison with the HL theory: 1—MT contribution $(d = 8.495$ Å); 2— *3D* AL contribution; 3—MT contribution $(d = 3.05$ Å).

As have been justified by experiments performed on YBCO films with different oxygen concentration [8,20] $\xi_c(0)$ can be written as

$$\xi_c(0) = d\varepsilon_0^{1/2}, \tag{5}$$

which gives a two times larger value of $\xi_c(0)$ than the HL theory. Besides, we have introduced the notation $[1.203(l/\xi_{ab})] = \beta$. As a result the parameter

$$\tau_\varphi \beta T=(\pi\hbar)/(8k_B\varepsilon_0) = A\varepsilon_0^{-1}, \tag{6}$$

where $A=(\pi\hbar)/(8k_B)$ = 2.988 $10^{-12}$ s, also becomes a function of $\varepsilon_0$ and can be calculated from the FLC analysis.

Expected MT-AL *(2D-3D)* crossover is clearly seen in Fig. 3 for $ln\varepsilon_0$ = –3.6. The fact enables us to determine $T_0$ and $\varepsilon_0$ with adequate accuracy. Using Eqs. (5, 6), and set $d$ = 8.495 Å, which is a dimension of the $SmFeAsO_{0.85}$ unit sell along *c*-axis [4,5], $\xi_c(0)$ = (1.4 ± 0.005) Å and $\tau_\varphi(100\ K)\beta = (11 \pm 0.03)\cdot 10^{-13}$s are derived from experiment. As expected, Eq. (2), with measured value of $\xi_c(0)$ and $C_{3D}$ = 0.083, fairly well describes the data just above $T_c^{mf}$ (dashed line 2). Thus, it was shown that, as well as YBCO cuprates, $SmFeAsO_{0.85}$ exhibits the *3D* fluctuation behavior close to $T_c$. Till now the $\sigma'(T)$ dependence is very similar to that observed for YBCO systems [8,20]. However the discrepancy appears when MT contribution is analysed. Really, substituting the measured values of $\xi_c(0)$ and $\tau_\varphi(100\ K)\beta$ into Eq. (3) we obtain curve 1 which apparently does not match the data. The finding suggests that our choice of $d$ is very likely wrong. To proceed with the analysis we suppose that $SmFeAsO_{1-x}$ becomes quasi-two-dimensional when $\xi_c(T)$, getting rise with temperature diminution, becomes at $T = T_{co} \approx 69$ K ($ln(\varepsilon_{c0}) \cong -1.55$) equal to $d$ = 3.05 Å, where (3.1÷3.0) Å is the distance between *As* layers in conducting *As–Fe–As* planes [4,21]. Below this temperature $\xi_c(T)$ is believed to couple the *As* layers by Josephson interaction. This approach gives the same value of $\xi_c(T) = d\varepsilon_{c0}^{1/2}$ = (1.4 ± 0.005) Å as calculated above using the crossover temperature $T_0$. We think this fact is to confirm our supposition. At the same time $\tau_\varphi(100\ K)\beta = (1.41 \pm 0.03)\cdot 10^{-13}$s is obtained in this case. Substitution of the measured value of $\xi_c(T)$ and this new $\tau_\varphi(100\ K)\beta$ into Eq. (3) enables us to fit the experimental data by the MT term just up to $T_{c0}$ (Fig. 3, curve 3). Thus, the fluctuation theories allow us to describe experiment in all temperature intervals of interest, which is about 15 K above $T_c$. The last finding suggests enhanced *2D* fluctuations in $SmFeAsO_{1-x}$.

An additional discrepancy is extremely small values of the *C*-factors ($C_{3D}$ = 0.083 and $C_{2D}$ = 0.082) in comparison with the YBCO films. The low values of *C*-factors are typical for YBCO films with defects [22]. However, in this case the MT fluctuation contribution is completely suppressed. On the contrary, the MT contribution observation just confirms the defect-free structure of the sample [20,22]. In $SmFeAsO_{1-x}$ the current flows mainly through FeAs conducting planes [1-5]. As a result, an effective volume of the sample could be many times lower than it's geometrical volume used to calculate $\rho(T)$. This evidently may results in very low values of $\sigma'(T)$ which in turn requires a small *C*-factors to fit the experiment.

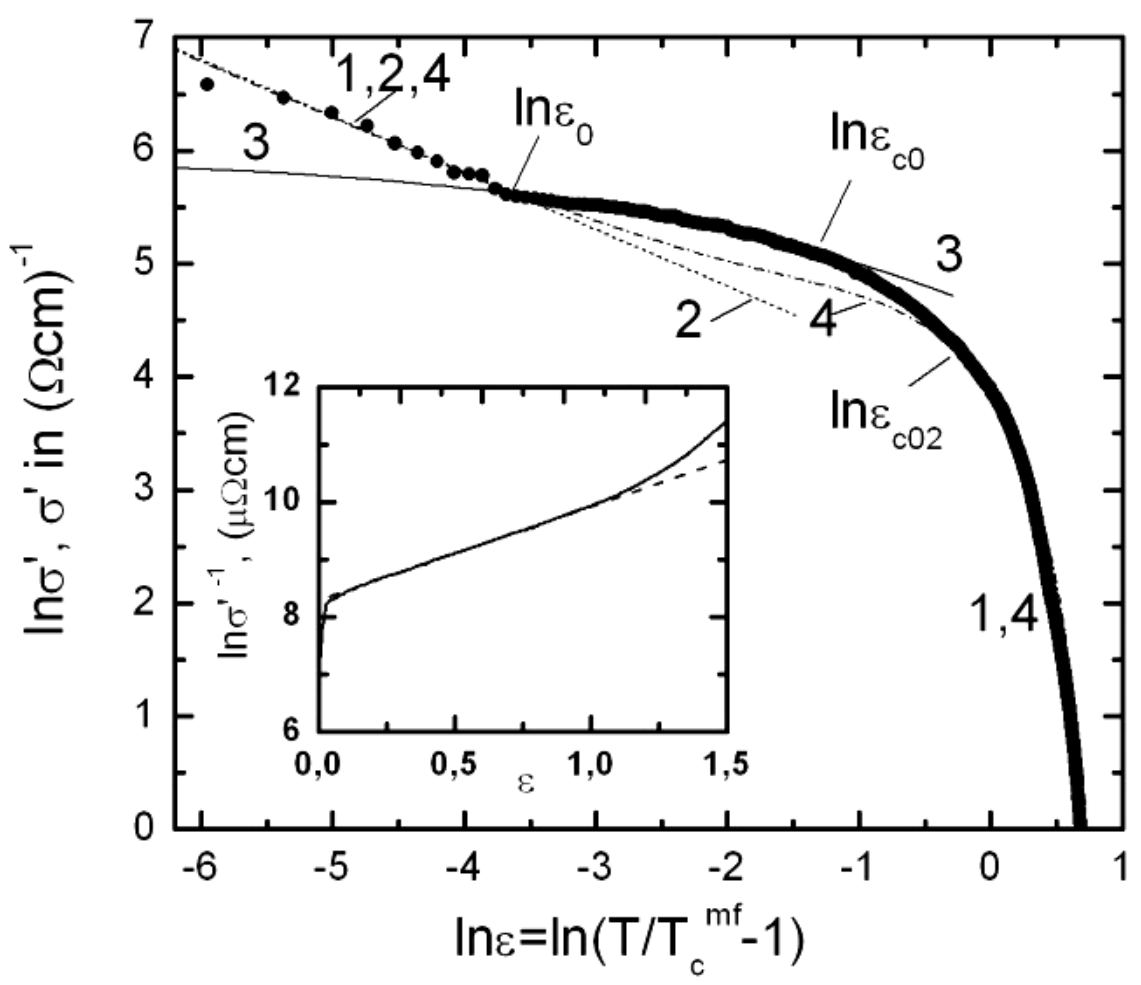

FIG. 4: $\sigma'(T)$ in the coordinates ln $\sigma'$ versus ln $\varepsilon$ (curve 1, dots) for sample $SmFeAsO_{0.85}$ in temperature interval from $T^*$ down to $T_c^{mf}$ in comparison with the theory: curve 2—*3D* AL contribution; 3 – *2D* MT contribution ($d$=3.05 Å), short dashed segment line 4 – Eq. (7). Inset: ln $\sigma'^{-1}$ versus $\varepsilon$ (solid line); dashed line − extrapolation of the rectilinear section.

### B. Pseudogap analysis

To analyse PG we assume that the excess conductivity σ'($T$) at $T_c^{mf} << T \leq T^*$ arises as a result of the formation of paired fermions (SBB or local pairs [12]) which satisfy the BEC theory [13-17,22,23]. Upon temperature diminution the local pairs transform into fluctuating Cooper pairs as T approaches $T_c^{mf}$ [7,8]. The conventional fluctuation theories (Fig. 4, curves 2,3) describe experiment up to $T_{c0} \approx$ 69 K only. To determine the temperature dependence of Δ*, referred to as a pseudogap, in the framework of our model it is necessary to describe the experimental dependence of $\sigma'(T)$ in the whole temperature interval from $T^* \approx 175\ K$ down to $T_c^{mf} \approx$ 57 K. The dynamics of pair-creation and pair-breaking above $T_c^{mf}$ must be taken into account in order to correctly describe experiment [7,13,24,25]. Ultimately, the equation for $\sigma'(\varepsilon)$ can be written as

$$\sigma'(\varepsilon) = A_4\,(1 - T/T^*)\,(\exp(-\Delta^*/T))\cdot$$

$$\cdot e^2/[16\hbar\xi_c(0)\sqrt{2\varepsilon_{c0}^*\sinh(2\varepsilon/\varepsilon_{c0}^*)}], \qquad (7)$$

where $A_4$ is a numerical factor which has the same meaning as $C$-factor in the FLC theory. Solving Eq. (7) for $\Delta^*$ we obtain

$$\Delta^*(T) = T\,\ln\{A_4\,(1 - T/T^*)/\,\sigma'(T)]\cdot$$

$$\cdot[e^2/[16\hbar\xi_c(0)\sqrt{2\varepsilon_{c0}^*\sinh(2\varepsilon/\varepsilon_{c0}^*)}]\}. \qquad (8)$$

Here $\sigma'(T)$ is the experimentally measured value of the excess conductivity (Fig.4, curve 1, dots) in the whole temperature interval from T* down to $T_c^{mf}$. All other parameters also come from experiment. As well as in YBCO compounds the reciprocal of the excess conductivity $\sigma'^{-1}(T)$ was found to be an exponential function of $\varepsilon$ between $ln(\varepsilon_{c0})$ and $ln(\varepsilon_{c02})$ ((69÷100)K). As a result, the function $ln\ \sigma'^{-1}$ versus $\varepsilon$, presented in the insets in Fig. 4, appears to be linear in this temperature range. As before, the slope $\alpha$ of this linear function determines the parameter $\varepsilon_{c0}^*$ used in both equations. Thus the only adjustable parameter remains coefficient $A_4$ which is chosen so that the curve, calculated with Eq. (7), fits the $\sigma'(\varepsilon)$ data in the region of $3D$ fluctuations near $T_c$ [7]. The curve constructed using Eq. (7) with the parameters $\varepsilon_{c0}^*$ = 0,616, $\xi_c(0)$ = 1,405Å, $T_c^{mf}$ = 56,99 K, $T^*$ = 175 K, $A_4$ = 1,98 and $\Delta^*(T_c)/k_B$ = 160 K is labelled with the number 4 in Fig. 4. It is seen, the curve 4 describes the experimental data well in the whole temperature interval of interest. The only exception is the $2D$ MT region where relatively small deviation down from experiment, negligible in the case of YBCO films [7], is observed. It is due to enhanced MT fluctuation contribution which reason has yet to be settled.

Now substituting the $\sigma'(T)$ data with the above set of parameters into Eq. (8) we obtain the expected dependences $\Delta^*(T)$ (Fig. 5). Unfortunately, the value of $\Delta^*(T_c)$ and in turn the ratio $2\Delta^*/k_BT_c$ in Fe-based superconductors remain uncertain. Reported in the literature values of $\Delta(0)$ and $2\Delta/T_c$ in $SmFeAsO_{1-x}$ range from $2\Delta \approx$ 37 meV ($2\Delta/T_c \sim$ 8, strongly coupled limit) obtained in measurements of far-infrared permittivity [26] down to $\Delta$ = (8 – 8.5) meV ($2\Delta/T_c \sim$ 3.55 – 3.8) measured by a scanning tunneling spectroscopy [2,27] which is very close to standard value 3.52 of the BCS theory (weakly coupled limit). It is believed at present that $SmFeAsO_{1-x}$ has two superconducting gaps $\Delta_1 \approx$ 6−7 meV and $\Delta_2 \approx$ 19−20 meV [5]. Besides we think that $\Delta^*(T_c) \sim \Delta(0)$ [29]. To feel more flexible, four curves are finally plotted in Fig. 5 with $\Delta^*(T_c)/k_B$ = 160 K ($2\Delta/T_c \sim$ 5.82), 140 K ($2\Delta/T_c \sim$ 5.0), 120 K ($2\Delta/T_c \sim$ 4.36) and 100 K ($2\Delta/T_c \sim$ 3.63) from top to bottom, respectively. Naturally, different values of coefficients $A_4$ correspond to each curve, whereas the other parameters remain unchangeable.

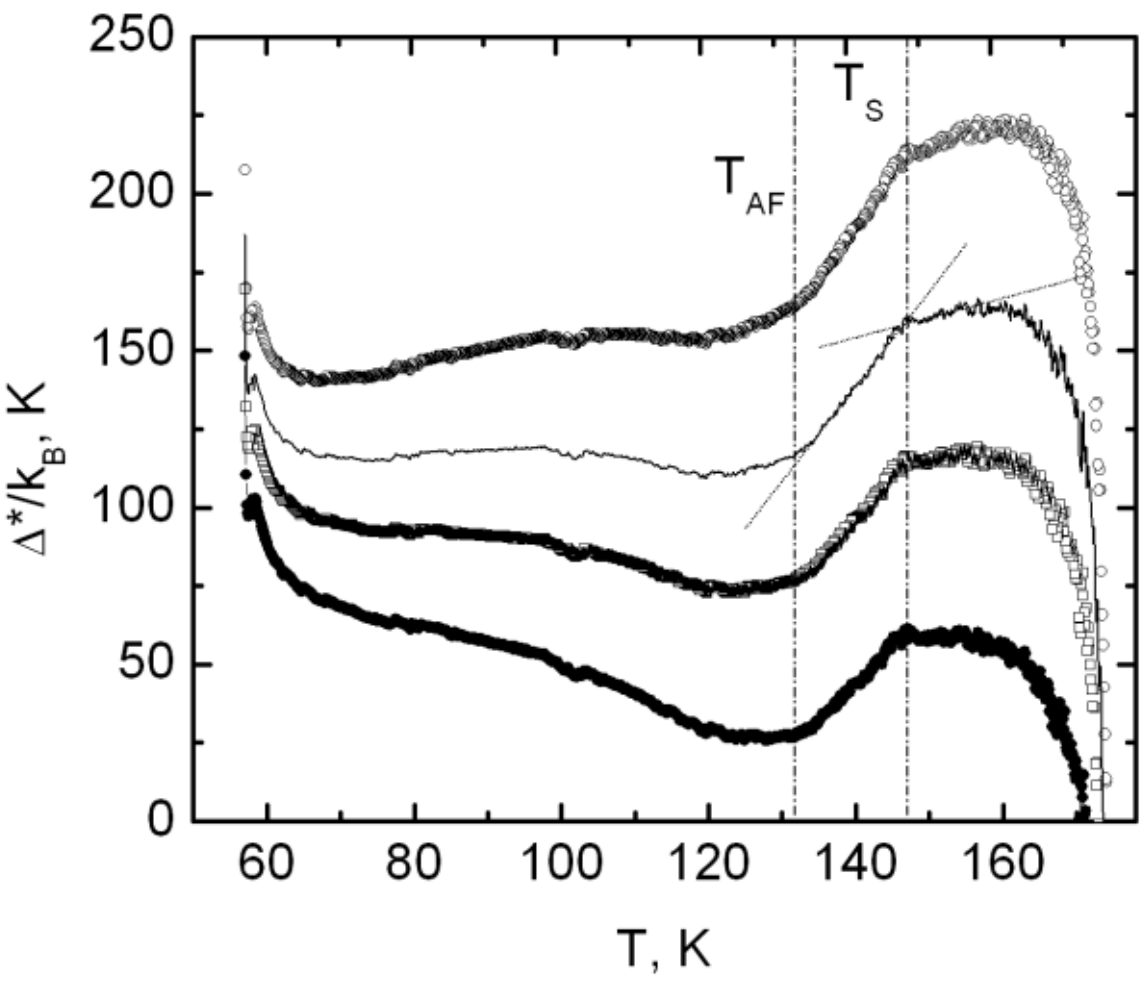

FIG. 5: $\Delta^*/k_B$ versus $T$ dependencies for $SmFeAsO_{0.85}$ with four different values of $\Delta^*(T_c)/k_B$ (see the text).

The most striking result is a sharp drop of $\Delta^*(T)$ at $T_S \sim$ 147 K as clearly illustrates the curve with $\Delta^*(T_c)/k_B$ =140 K plotted without symbols. At $T_S$ there occurs a structural transition in SmFeAsO [21] expected to be a transition to spin-density wave (SDW) regime in Sm-based compounds too [5,19]. Below $T_S$ (Fig. 5) $\Delta^*(T)$ linearly drops down to $T_{AF} \approx$ 133 K which is attributed to AF ordering of Fe spins in SmFeAsO [3,30]. Below $T_{AF}$ the $\Delta^*(T)$ behaviour evidently depends on the $\Delta^*(T_c)$ value (Fig. 5). Strictly speaking it is difficult to say at present is $T_{AF}$ = $T_N$ of the whole system or not because the AF ordering of Sm spins occurs at ~ 5 K only [3-5,30].

Found $\Delta^*(T)$ behaviour is in a good agreement with the theory by Machida, Nokura and Matsubara (MNM) developed for AF superconductors in which the AF ordering may coexist with superconductivity [31]. In accordance with the MNM theory at $T_N < T_c$ such a system undergoes a transition in SDW regime due to the formation of AF energy gap on the Fermi surface. The SDW ordering has to suppress the order parameter *Δ(T)*. Predicted *Δ(T)* decrease at $T_N$ similar to that shown in Fig. 5 was recently observed in AF superconductor $ErNi_2B_2C$ with $T_c \approx 11$ K and $T_N \approx 6$ K below which the SDW ordering occurs in the system [32]. Important in our case is the fact that we see the peculiarities of *Δ*(T)* in the PG state i.e. well above $T_c$ and, besides, in the sample with approximately optimal oxygen concentration in which no peculiarities at $T_S$ and $T_{AF}$ are expected [3-5]. In this connection it is worth to emphasize that recent phase diagrams allow the overlapping of SDW and SC state in $SmFeAsO_{1-x}F_x$ up to x ≈ 0.15 [21,33].

## IV. CONCLUSION

A systematically study of the excess conductivity *σ′(T)* derived from the resistivity measurements of the superconductor $SmFeAsO_{0.85}$ was performed. Obtained results have apparently indicated the applicability of our approach developed for YBCO HTS's [7,8] to the analysis of the FLC and PG in iron-based superconductors, at least in $SmFeAsO_{1-x}$ systems. The $ln\sigma'^{-1}$ versus $ln\varepsilon$ dependence (Fig. 3,4) very similar to that obtained for optimally doped YBCO films was found. The exception is enhanced MT fluctuation contribution required *d* to be ~ (3.05 ± 0.5) Å which is most likely the distance between *As* atoms in conducting *As-Fe-As* plane, and extremely small values of scaling C-factors. The reason for that has yet to be settled.

At the same time the rather specific temperature dependence of the PG *Δ*(T)* was observed (Fig. 5). The more striking result demonstrated by all curves independently on $\Delta^*(T_c)/k_B$ value is the pronounced reduction of *Δ*(T)* at $T_S \sim 147$ K where transition in SDW regime likely occurs. Interestingly, no evident peculiarities on the *σ′(T)* (Fig. 4) in this temperature region is observed. Below $T_S$ *Δ*(T)* is linear down to $T_{AF} \sim 133$ K which is attributed to antiferromagnetic ordering of Fe spins [3-5,21,33]. Note that no such peculiarities of *Δ(T)* in superconducting state of $SmFeAsO_{1-x}F_x$ is observed [34]. The result confirms the current conclusion as for SDW ordering absence in SC state of the Fe-based compounds [3-5]. Found *Δ*(T)* diminution can be explained by the MNM theory [30] which predicts the suppression of the order parameter by the SDW ordering. But we have to emphasize that we observe the *Δ*(T)* reduction in the PG state i.e. well above $T_c$. The finding suggests the presence of paired fermions, characterized by *Δ*(T)*, in $SmFeAsO_{0.85}$ in the PG region most likely in the form of SBB. Thus, the SBB presence seems to be the common feature of the PG formation both in cuprates and Fe-based HTS's.

However, compare with YBCO systems the *Δ*(T)* behaviour found for $SmFeAsO_{1-x}$ turned out to be rather intrigueous since no peculiarities in the normal state of the Sm-based compounds with optimal oxygen concentration are expected [3-5]. Thus, the finding suggests a conclusion that the phase diagrams reported for Sm-based superconductors [3-5] are not completely established. The diagrams turned out to be more complicated than those reported for $LaFeAsO_{1-x}F_x$ [35], $CeFeAsO_{1-x}F_x$ [36] and actually for many others Fe-based compounds [3-5]. Evident SDW regime existence in the normal state is reported for $SmFeAsO_{1-x}F_x$ with x up to at least ~ 0.15 in Ref. [21]. And finally, much more complicated phase diagram taking account of magnetic subsystem complexity in $SmFeAsO_{1-x}F_x$ and being to a certain extent in more agreement with our experimental results was reported in Ref.[33]. Nevertheless, more experimental results are evidently required to clarify the issue.

**Acknowledgments**

We thank Yu. G. Naidyuk for useful discussions of the results obtained in the study of $ErNi_2B_2C$.

* Electronic address: solovjov@ilt.kharkov.ua